\documentclass[12pt,a4paper]{article}

\usepackage{ifpdf}

\usepackage{amsmath}
\usepackage{amssymb}
\usepackage{amsfonts}
\usepackage{graphicx}
\usepackage[english]{babel}
\usepackage{palatino}
\usepackage[T1]{fontenc}
\usepackage{fancyhdr}
\usepackage{color}
\usepackage[flushleft]{threeparttable}
\usepackage{url}

\usepackage{etoolbox}
\let\bbordermatrix\bordermatrix
\patchcmd{\bbordermatrix}{8.75}{4.75}{}{}
\patchcmd{\bbordermatrix}{\left(}{\left[}{}{}
\patchcmd{\bbordermatrix}{\right)}{\right]}{}{}

\makeatletter

\renewcommand\@biblabel[1]{} \makeatother

\usepackage[top=2cm, bottom=2cm, outer=2cm, inner=2cm]{geometry}
\usepackage{xargs}
\usepackage[dvipsnames]{xcolor}
\usepackage[colorinlistoftodos,prependcaption,textsize=tiny]{todonotes}
\usepackage[authoryear,round]{natbib}

\usepackage{array}
\newcolumntype{H}{>{\setbox0=\hbox\bgroup}c<{\egroup}@{}}

\usepackage{tabu}

\DeclareMathOperator{\tr}{tr}

\begin{document}

\title{Occupational Mobility: Theory and Estimation for Italy\thanks{This research has benefited from the financial support of the Italian Ministry of University and Research as part of the PRIN 2009 programme (grant protocol number 2009H8WPX5). The authors would like to thank seminar participants at University of Pisa, La Sapienza, University of Rome, and, in particular, Fulvio Corsi, Maurizio Franzini and Michele Raitano for their very useful comments.}}

\author{Irene Brunetti\thanks{Irene Brunetti, INAPP - National Institute for Public Policy Analysis, e-mail: i.brunetti@inapp.org.} - Davide Fiaschi\thanks{Davide Fiaschi, Dipartimento di Economia e Management, University of Pisa, e-mail: davide.fiaschi@unipi.it.}}

\date{\today}

\maketitle

\begin{abstract}
This paper presents a model where intergenerational occupational mobility is the joint outcome of three main determinants: income incentives, equality of opportunity and changes in the composition of occupations. The model rationalizes the use of transition matrices to measure mobility, which allows for the identification of asymmetric mobility patterns and for the formulation of a specific mobility index for each determinant. Italian children born in 1940-1951 had a lower mobility with respect to those born after 1965. The steady mobility for children born after 1965, however, covers a lower structural mobility in favour of upper-middle classes and a higher downward mobility from upper-middle classes. Equality of opportunity was far from the perfection but steady for those born after 1965. Changes in income incentives instead played a major role, leading to a higher downward mobility from upper-middle classes and lower upward mobility from the lower class.
\end{abstract}



\newpage

\section{Introduction}

The downward trend in intergenerational mobility, i.e. the increasing correlation between parents and children's socio-economic status, is a very debated issue (see, e.g., \citealp{Corak2013}, \citealp{DobbsEtAl2016}, and \citealp{theEconomist2020}). Mobility is at the root of the contemporary society, and already in 1821 Thomas Jefferson in The Founders' Constitution argued: ``The transmission of this property from generation to generation, in the same name, raised up a distinct set of families, who, being privileged by law in the perpetuation of their wealth were thus formed into a Patrician order [$\ldots$]. From this order, too, the king habitually selected his counsellors of State [$\ldots$] To annul this privilege, and instead of an aristocracy of wealth of more harm and danger than benefit to society, to make an opening for the aristocracy of virtue and talent, which nature has wisely provided for the direction of the interests of society, and scattered with equal hand through all its conditions, was deemed essential to a well-ordered republic'' (\citealp{Jefferson1944}). 
In a more economic (and contemporary) jargon, a society with a high mobility is seen as both more \textit{just} in terms of equality of opportunity, that is initial positions and not individual efforts decide individual achievements (\citealp{Roemer1998}), and more \textit{efficient}, i.e. able to attain ``the largest possible value of advantages given a production possibility set'' (\citealp{BourguignonEtAl2007B}). 

However, a multifaceted phenomenon as intergenerational mobility prevents the formulation of an unifying framework. This paper focuses on (absolute) \textit{intergenerational occupational mobility}, providing a micro-foundation of transition probabilities between occupational status of parents and children, and then an estimate of them on a sample of Italian heads of households born in the period 1940-1977. In particular, the model considers three main determinants of overall mobility: \textit{income incentives}, or rather the choice based on differential returns of occupations; \textit{equality of opportunity}, that is occupation possibility set available to each individual (together they concur to determine \textit{true mobility}); and \textit{occupational shifts}, i.e. the change in the composition of occupations, which determines \textit{structural mobility} (\citealp{Prais1955}).

The model rationalizes the use of transition matrices to measure mobility. The micro-foundation of the elements of these matrices, providing a detailed picture of intergenerational mobility among occupations, allows for: i) the identification of \textit{asymmetric patterns} across the distribution of occupations (e.g., of the higher downward mobility at top than at the middle of distribution), and for subgroup comparisons across the entire distribution (\citealp{FieldsOk1999} and \citealp{JanttiEtAl2006}); and ii) the estimation of specific \textit{mobility indexes} measuring income incentives, equality of opportunity, and occupational shifts (in the spirit of \citealp{Shorrocks1978}). 

The choice to focus on occupation instead of income, wealth or earnings, is based on several considerations. Apart from the common unavailability of data on parental income/wealth, occupational classes more properly encompass several key aspects of socio-economic characteristics of individual's family background, such as their position in the social scale, and better reflect the socio-economic status of individuals, as their social prestige and the control on their own life (\citealp{GanzeboomTreiman1996}, \citealp{Barone2012}, and \citealp{RaitanoVona2015}). Moreover, occupational classes allow to take into account the shift in the composition of occupations, which \citet{Prais1955}, \citet{Breen2004} and \citet{LongFerrie2013} claim as a key explanatory factor of mobility. Finally, \citet{EriksonGoldthorpe2002} show that occupational class is less affected by measurement error and it is a better predictor of life chances and lifetime earnings when compared to snapshot measures of income (see also \citealp{Atkinson1980}).\footnote{\citet{BjorklundJantti2000} summarize some of the relative merits of the use of occupational classes to estimate intergenerational mobility, and discuss scenarios in which it provides different results from those where intergenerational mobility is estimated by individual income.}

Our empirical analysis points out that Italian children born in 1940-1951 had a lower mobility with respect to those born in 1951-1965, while children born in 1965-1977 experienced the same mobility. This steady mobility however covers a lower structural mobility in favour of upper-middle classes and a higher downward true mobility from upper-middle classes. 
Looking at the determinants of true mobility we find that the equality of opportunity, far from the perfection, increased over time for children with father in lower-middle classes and decreased for children with father in upper class. Overall, equality of opportunity increased over time for children born before 1965, and become steady for those born after.
On the contrary, changes in income incentives notably contributed to the increased true mobility. However, this outcome covers two opposite tendencies: the higher income incentives for 
children with father in lower class to remain in the same class (i.e. a lower upward mobility) as opposed to the lower income incentives for children with father in upper-middle classes to remain in the same class (i.e. a higher downward mobility), with the latter effect strongly prevailing on the first.
The observed changes over time in differential returns and riskiness of occupations, together with anecdotal evidence of the increasing cost of education in Italy, could contribute to explain our estimated changes in income incentives.

This paper contributes both to the theoretical and empirical literature on mobility. As regards theory, we provide a model which jointly considers three possible determinants of occupational mobility: income incentives, equality of opportunity and occupational shifts and relate these to transition matrices across occupational classes. The first two determinants are commonly present in the literature; in particular, \citet{DardanoniEtAl2006}) qualify the key concepts of circumstances and effort as proposed by \cite{Roemer1998} for the measurement of the equality of opportunity, and discuss four channels through which circumstances affect income opportunity. Occupational shifts are instead generally neglected in the recent literature, despite the actual deep structural change experimented by many countries. 

More importantly, the micro-foundation of transition matrices allows to circumvent some key limitations in the identification of true mobility and equality of opportunity generally due to the lack of data. \citet{RoemerTrannoy2016} discuss how the availability of rich versus poor datasets leads to two alternative approaches to the estimate of equality of opportunities. Within the case of a rich dataset, \citet{BourguignonEtAl2007A} propose an econometric strategy based on a system of simultaneous equations describing the process of attainment of the socio-economic status of individuals, and the impact of circumstances, and other control variables, on individual effort. Differently, within the case of poor dataset, \citet{ChecchiPeragine2010} propose a theoretical methodology to measure the equality of opportunities based on a decomposition of the mean logarithmic deviation index of inequality where opportunity inequality is obtained as difference between total income inequality and effort inequality. However, both approaches relies on restrictive identification assumptions: for example, \citet{ChecchiPeragine2010} assume that any variation in individual income is to be ascribed to individual effort and opportunities are proxied by parents' education. 
In our theoretical framework, with an information set corresponding to the case of poor dataset, the constraints directly derived by the theoretical model are sufficient to estimate the equality of opportunity and also income incentives.

As regards the empirical literature, we provide the most recent estimates on true and structural occupational mobility in Italy. We therefore complement the analysis by \citet{SchizzerottoCobalti1994} and \citet{Pisati2000} considering individuals born between 1920-1970 in Italy. More importantly, applying our methodology for the identification of the determinants of Italian mobility, we find results similar to \citet{ChecchiDardanoni2003}, i.e. a lower structural mobility and a higher true mobility over time. With respect to \citet{ChecchiPeragine2010} and \citet{BrunoriEtAl2013}, who find that equality of opportunity explains from 20\% to 73\% ex-post Italian (income) inequality, we estimate that, limiting to true mobility, only about 43\% of opportunities are exploited in the most recent cohort of children.

The paper is organized as follows. Section \ref{sec:litReview} contains a literature review and some clarification on the concept of mobility used in our theoretical model; Section \ref{sec:MobilityModel} presents the model and discusses some mobility indexes; Section \ref{sec:EmpiricalAnalysis} is devoted to the empirical analysis on Italy. Section \ref{sec:ConcludingRemarks} concludes.

\section{Literature Review \label{sec:litReview}}

Intergenerational mobility refers to changes in \textit{socio-economic status} from one generation to another, in particular to the relationship between the status of parents and of their children (\citealp{FieldsOk1999}). Intergenerational mobility is a key indicator of ``openness'' of a society or of its equality of opportunity (\citealp{Atkinson1980} and \citealp{KanburStiglitz2016}). In this regard, \citet{Roemer1998} argues that a mobile society is more just flatting the ``playing field'' among individuals competing for the same position, minimizing the importance of ``circumstances'', that is of the variables that are beyond the individual's control, and maximizing the role of individual effort in the individual achievements (\citealp{DardanoniEtAl2006}). Mobility is also related to \textit{meritocracy}, a term introduced by \cite{Young1958}, and largely used in literature and in public debate to indicate a socio-economic system where the merit, the sum of individual effort and talent, is the only factor deciding the social status of an individual.

\citet{Conlisk1974} represents the seminal theoretical paper on the transmission across generations of individual socio-economic status, as measured by income or earnings. \citet{BeckerTomes1979} and \citet{BeckerTomes1986} proposes a model where mobility is the result of parents' invest in the education of their children to maximize their expected utility. They show that the correlation between parents and children's economic status positively depends on the degree of inheritability of genetic and cultural attributes and on the parents' propensity to leave a bequest and to invest in education. In their optimization framework a provision of public education could have no effect on mobility since a higher public investment in education is offset by a lower private investment. \citet{Goldberger1989}, however, proposes a detailed theoretical argument against this conclusion. In the same line, \citet{Solon2004} finds that mobility decreases with the efficacy of private investment and the progressiveness of public investment in children's education. \citet{Durlauf1996} proposes a model where parents decide the income of their children through the choice of the place of living and of the amount of investment in education. Local financing of education, as well as some sociological factors, incentives wealthier families to segregate themselves into economically homogeneous neighbourhoods and this segregation induces persistent inequality, poverty, and low socio-economic mobility. \citet{Piketty2000} reviews different theoretical models on persistent inequality and intergenerational mobility. He discusses how persistent inequality is explained either i) with the presence of imperfect capital markets, which impedes talented individuals with poor parents to efficiently invest; or ii) with the combination of direct family transmission of productive abilities and ambitions, and efficient human capital investments. While in case i) a lower economic mobility implies a lower efficiency of economy, in case ii) a lower mobility would have only implications for distributive justice considerations.
In this regard, \citet{Loury1981} has firstly introduced the presence of credit constraints into a \citet{BeckerTomes1979}-type model of mobility. \citet{BanerjeeNewman1993} discuss how credit constraints and individual endowments can crucially affect occupational choice of individuals. Finally, in a more recent paper, \citet{AbbottGallipoli2017} show the importance of supply side introducing into the Becker-Tomes model a sector in which workers' human capital are complements; the return to parental human capital investments is therefore lower where skill complementarity is more intense, and this results in a higher intergenerational mobility.

Looking at empirical studies, \citet{Solon1992} and \citet{Zimmerman1992} evaluate the mobility in US on the base of the estimate of intergenerational earnings elasticities (IGEs). After correcting for the measurement error in the reported individual earnings, they show that the US society appears much less mobile than most previous studies have found. \citet{Piraino2007} and \citet{Mocetti2007} show that the IGE in Italy, estimated at a level of about 0.5, is lower than in US. \citet{Bartholomew1973} and \citet{Shorrocks1978} propose an alternative approach to the measurement of mobility based on transition matrices and synthetic mobility indexes arguing that they are more informative than the estimate of IGEs. Using this approach, \citet{ChecchiEtAl1999} find that Italy has a lower level of mobility in education, income and occupations than US. \citet{JanttiEtAl2006} estimate both IGEs and transition matrices for US, the United Kingdom, Denmark, Finland, Norway and Sweden, finding significantly lower rates of upward mobility from the bottom of the distribution in US compared to the Nordic countries. Finally, \citet{Corak2016} reports that US recently display one of the the lowest economic mobility among developed countries; in particular, current intergenerational earnings mobility in US is almost three times lower than in Canada.

Less attention in literature is devoted to the \textit{empirical identification} of the determinants of intergenerational mobility, which instead appears to be crucial to design an effective social policy (\citealp{KourtellosEtAl2016}). \citet{Goldberger1989} discusses how Becker-Tomes framework is crucially affected by the lack of identification of different factors at the root of socio-economic mobility. \citet{ChettyEtAl2014} remark how the empirical identification is strongly limited by the lack of a common theoretical framework which can guide the choice of variables to be included in the analysis. \citet{Sorokin1927}, \citet{BlauDuncan1967} and \citet{Boudon1973} argue that social mobility is the joint outcome of several variables which interact among themselves: social origin (family background, school, etc.), individual's motivations, differential demographic dynamics among different strata of population (birth, death, in-and out- migration) and shifts in the occupational structure (see \citealp{Prais1955} and \citealp{Thurow1976}). In particular, \citet{Sorokin1927} explains how the effects of such variables on mobility do not add up, but they act in combination among themselves. In addition, \citet{Prais1955} argues that the observed occupational mobility is the outcome of two distinct phenomena: the choices of individuals (Prais indicates it as the "true" occupational mobility), and the changes in the occupational structure and/or in the reproduction (net fertility) rates of classes. The same distinction is put forth by another strand literature on mobility, which divides the total observed mobility between the \textit{structural} and the \textit{exchange} mobility (see, e.g., \citealp{VanKerm2004}). The former is the amount of mobility generated by the fact that the occupations distribution of children differs from the distribution of their fathers; the latter is defined as the part of the total mobility which is not structural (\citealp{Boudon1973}). \citet{ChecchiDardanoni2003} compare several mobility indexes in terms of their capacity to effectively measure structural and exchange mobility. 

Among the possible factors affecting mobility, \citet{Fox2016} shows that wealthier families having greater educational access, greater occupational networks, and more neighbourhood choices, promote better child outcomes. \citet{OlivettiPaserman2015} argue that the declining trend of mobility could be explained by the drop in fertility, the increasing inequality, and the raising return to human capital (see \citealp{Solon2004} and \citealp{Corak2016}). \citet{FerreiraEtAl2017} investigate the relationship between inequality of opportunity, driven by circumstances at birth, and economic growth. Their results point out that, while both income inequality and inequality of opportunity are negatively associated with growth, there is not any statistically significant relationship between inequality due to circumstances and growth.

As regards social mobility in Italy, \citet{Checchi2003} shows that educational choices strongly depend on parents' education and only marginally on income incentives and/or liquidity constraints. \citet{DiPietroUrwin2003} find that capital market imperfections affected investment in human capital among the poor Italian households; however, after controlling for this effect, children with father in lower-middle class still continue to show less opportunities to belong to a higher occupational class. For Italian heads of households born in the period 1930-1970, \citet{ChecchiDardanoni2003} find that structural mobility has been declining, while changes in the openness of the society increased true mobility. 

\section{A Model of Occupational Mobility  \label{sec:MobilityModel}}

Inspired by literature review in Section \ref{sec:litReview}, our model of intergenerational occupational mobility encompasses three key determinants of occupational mobility, i.e.: i) the \textit{income and social incentives} of different occupations (see, e.g., \citealt{Corak2013}); ii) the \textit{opportunities} deriving from  family cultural and genetic attributes and socio-economic environment (the concept of ``inheritated endowments'' proposed in \citealp{BeckerTomes1986} and in Chapter 3 of \citealp{Corak2004}); and, finally, iii) the \textit{occupational shifts}, i.e. the change in occupational possibility set determined by the production side of economy (\citealp{Prais1955} and \citealp{Thurow1976}).
In particular, occupational status attainment is associated both to \textit{individual's motivations} and to individual \textit{social origin} (\citealp{BlauDuncan1967}).
Individual motivations depend on the characteristics of occupations, as its associated income and social prestige, while social origin, which \citet{DardanoniEtAl2006} include on the more wider concept of ``circumstances'', and we denote as \textit{opportunities}, can be identified as: ``[$\dots$] the aspects of the environments of individuals that affect their achievement of the objective, and for which the society in question, or the policymaker, does not wish to hold individuals responsible.'' In the following, we will denote \textit{true mobility} the mobility due to both \textit{income incentives} and \textit{equality of opportunity}, and \textit{structural mobility} the mobility due to \textit{occupational shifts} (the same terminology used in \citealp{ChecchiDardanoni2003} and \citealp{VanKerm2004}).

Consider an economy with only three classes of occupations: \textit{Working} class (\textbf{Wc}), \textit{Middle} class (\textbf{Mc}) and \textit{Upper} class (\textbf{Uc}). This partition of society has a long tradition, starting from the rise of Middle Class from the end of the 19th century in the most of developed countries, and it is well discussed in the literature on social classes (\citealp{GoldthorpeHope1974}). We will postpone to Section \ref{sec:EmpiricalAnalysis} how this partition is very effective to represent the actual Italian occupational mobility.
 
Assume that the (log of) life-time (indirect) utility of individual $i$, $U_{i}$, is given by:
\begin{equation}
 \log U_{i}= V_i + \epsilon_i,
 \label{eq:lifetimeRandomUtility}
\end{equation}
where $V_i$ is the the systematic component of utility, and $\epsilon_i$ the random component.
Both components depends on the occupational class of individual $i$. In particular, as the systematic component we assume that:
\begin{equation}
V_i = \left\{	 \begin{array}{ccc}
\mu_W & \text{if } i \in \text{Wc}; \\
2 \theta_i \mu_M & \text{if } i \in \text{Mc}; \text{ and} \\
2 \theta_i \mu_U & \text{if } i \in \text{Uc}. \\
\end{array}
\right.
\label{eq:specificationSystematicComponent}
\end{equation}
Parameter $\theta_i \in [0,1]$ represents ``the endowment'' of individuals in \citet[S4]{BeckerTomes1986}, i.e. ``cultural and genetic attributes [...] trasmitted from parents to children''; or ``the child's luck variable'' as in \citet[507]{Goldberger1989}: ``the capital that child receives automatically and effortlessly from his parents-genes, reputation, culture, learning, skill and goals that his parents provided at no cost''.
At the time of the choice of occupational class $\theta_{i}$ is assumed to be known by individual $i$, but it has a stochastic nature (see below for more details).
As regards the stochastic component, assume that:
\begin{equation}
\epsilon_i \sim N\left(0,\sigma^2_i\right),
\label{eq:errorComponent}
\end{equation}
where:
\begin{equation}
\sigma^2_i = \left\{	 \begin{array}{ccc}
\sigma^2_W & \text{if } i \in \text{Wc}; \\
\sigma^2_M & \text{if } i \in \text{Mc}; \text{ and} \\
\sigma^2_U & \text{if } i \in \text{Uc}. \\
\end{array}
\right.
\label{eq:specificationVarianceErrorComponent}
\end{equation}

In order to reflect the different life-time utility associated to the three classes assume:
\begin{equation}
\mu_W \leq \mu_M \leq \mu_U,
\label{eq:assumptionDifferentExpectedUtilityDifferentOccClass}
\end{equation}
while no restrictions are imposed on the value of $\sigma^2_W$, $\sigma^2_M$, and $\sigma^2_U$.
From one hand, the dispersion of life-time utility is expected to be higher in Uc than in Mc and Wc for the importance of individual attributes on individual income; on the other hand, the types of occupations in Wc, i.e. low-skilled jobs, make individuals in this class more subject to unemployment risk.

Taking together Eqq. (\ref{eq:lifetimeRandomUtility})-(\ref{eq:specificationVarianceErrorComponent}) aims at representing the substantial differences in the three occupational classes: life-time utility of individuals belonging to Wc is assumed to be independent of individual attributes and lower than the one of Mc and Uc. Wc should ideally include the most of blue-collars and workers in low-skilled jobs. On the contrary, an individual belonging to Mc and Uc should exploit her specific attributes to increase her lifetime utility; Mc and Uc should therefore include all individuals undertaking an entrepreneurial and/or professional activity based on own personal abilities and/or on education. In this respect, in order to take into account that the most of occupations in Mc and Uc require an adequate level of skills, assume that the access to Mc and Uc requires to bear a cost $\tilde{c}_e$ (expressed in units of utility). Moreover, since such skills can be acquired by formal education and/or by benefiting of a favourable social environment, we allow that $\tilde{c}_e$ can negatively depend on individual attributes $\theta_i$.\footnote{A cost of education negatively depending on individual talent is generally assumed in the theory of signalling (see, e.g., \citealp{Weiss1995}).} In particular:
\begin{equation}
\tilde{c}_M^e = \left(1-\delta \theta_i \right){c}^M_e,
\label{eq:costAcquiringSkillsMediumClass}
\end{equation}
is the cost to bear to access to Mc, and
\begin{equation}
\tilde{c}_U^e = \left(1-\delta \theta_i \right){c}^U_e,
\label{eq:costAcquiringSkillsCapitalistClass}
\end{equation}
is the cost to bear to access to Uc, where $\delta \in \left[0,1\right]$ measures the magnitude of negative impact of individual attributes on the cost of meeting the required skills to access to Mc and Uc.
In order to reflect the hierarchy among different occupational classes we assume that:
\begin{equation}
{c}^U_e > {c}^M_e.
\label{assunzione:costoEducation}
\end{equation}

\subsection{Income Incentives}

Individual $i$ prefers to belong to Mc instead of Wc if and only if her net expected utility in Mc is higher than in Wc (with a slight abuse of notation):
\begin{equation}
\text{Mc} \succsim \text{Wc} \Longleftrightarrow 
\mathbf{E}\left[\log U_M\right] - \tilde{c}_M^e \geq\mathbf{E}\left[\log U_W\right] + \sigma_{WM}^{RP},
\label{eq:IncentiveAssumption}
\end{equation}
where $\sigma_{WM}^{RP}$ is the \textit{risk premium} of belonging to Wc instead of Mc. We assume that individual $i$ is risk-adverse or risk-neutral, which implies that:
\begin{equation}
\sigma_{WM}^{RP} = \sigma^{RP}\left(\dfrac{\sigma_{M}^2}{\sigma_{W}^2}\right) \text{, with }  \dfrac{\partial \sigma^{RP} }{\partial \sigma_{M}^2 / \sigma_{W}^2} \geq 0 \text{ and }  \sigma^{RP}\left(1\right)=0.
\label{eq:riskPremium}
\end{equation}
Given Eqq. (\ref{eq:lifetimeRandomUtility}), (\ref{eq:errorComponent}), (\ref{eq:costAcquiringSkillsMediumClass}) and (\ref{eq:riskPremium}), Condition (\ref{eq:IncentiveAssumption}) can be written as:
\begin{eqnarray} 
\text{Mc} \succsim \text{Wc} \Longleftrightarrow \theta_{i} \geq \lambda_M  \equiv \underbrace{\dfrac{\mu_{W}}{2\mu_{M} + \delta c_M^e}}_{\text{Return premium}} + \underbrace{\dfrac{\sigma^{RP}\left(\sigma_{M}^2/\sigma_{W}^2\right)}{2\mu_{M}+ \delta c_M^e}}_{\text{Risk premium}} + \underbrace{\dfrac{c_M^e}{2\mu_{M} +\delta c_M^e}}_{\text{Cost of access}}.
\label{eq:ConditionIncomeIncentives}
\end{eqnarray}

On the base of (individual) \textbf{income incentives} a high level of $\lambda_M$ means a \textit{lower} probability for individual $i$ to belong to Mc (we use ``probability'' because $\theta_i$ has a random component, see below); at \textit{aggregate level} $\lambda_M$ instead measures the probability mass of individuals that decide to belong to Wc. Eq. (\ref{eq:ConditionIncomeIncentives}) indicates that the mass of individuals in Mc negatively depends on i) $\mu_{W}/\mu_{M}$, the inverse of expected occupational \textit{return premium} to belong to Mc; ii) $\sigma_{M}^2/2\sigma_{W}^2$, the \textit{risk premium} to belong to Mc; iii) $c_M^e$ the \textit{cost of meeting the required skills to access} to Mc; and iv) the magnitude of negative \textit{impact of individual attributes on the cost of meeting the required skills to access} to Mc.

In the same fashion, individual $i$ prefers to belong to Uc instead of Mc if and only if her net expected life-time utility in Uc is higher than in Mc, i.e.:
\begin{equation}
\text{Uc} \succsim \text{Mc} \Longleftrightarrow 
\mathbf{E}\left[\log U_U\right] - \tilde{c}_U^e \geq\mathbf{E}\left[\log U_M\right] + \sigma^{RP}\left(\dfrac{\sigma_{U}^2}{\sigma_{W}^2}\right),
\label{eq:IncentiveAssumptionUc}
\end{equation}
which, in light of Eqq. (\ref{eq:lifetimeRandomUtility}), (\ref{eq:errorComponent}), (\ref{eq:costAcquiringSkillsMediumClass}) and (\ref{eq:riskPremium}), can be expressed as:
\begin{eqnarray} 
\text{Uc} \succsim \text{Mc} \Longleftrightarrow \theta_{i} \geq \lambda_U  \equiv \dfrac{\mu_{M}}{2\mu_{U} + \delta c_U^e} + \dfrac{\sigma^{RP}\left( \sigma_{U}^2 / \sigma_{M}^2\right)}{2\mu_{U}+ \delta c_U^e} + \dfrac{c_U^e}{2\mu_{U} +\delta c_U^e}.
\label{eq:ConditionIncomeIncentivesUc}
\end{eqnarray}

Finally, individual $i$ prefers to belong to Uc instead of Wc if and only if her net expected utility in Uc is higher than in Wc, i.e.:
\begin{equation}
\text{Uc} \succsim \text{Wc} \Longleftrightarrow 
\mathbf{E}\left[\log U_U\right] - \tilde{c}_U^e \geq\mathbf{E}\left[\log U_W\right] + \sigma^{RP}\left(\dfrac{\sigma_{U}^2}{\sigma_{W}^2}\right),
\label{eq:IncentiveAssumptionUcVsWc}
\end{equation}
which, in light of  Eqq. (\ref{eq:lifetimeRandomUtility}), (\ref{eq:errorComponent}), (\ref{eq:costAcquiringSkillsMediumClass}) and (\ref{eq:riskPremium}), can be expressed as:
\begin{eqnarray} 
\text{Uc} \succsim \text{Wc} \Longleftrightarrow \theta_{i} \geq \lambda^W_U  \equiv \dfrac{\mu_{W}}{2\mu_{U} + \delta c_U^e} + \dfrac{\sigma^{RP}\left( \sigma_{U}^2 / \sigma_{W}^2\right)}{2\mu_{U}+ \delta c_U^e} + \dfrac{c_U^e}{2\mu_{U} +\delta c_U^e}.
\label{eq:ConditionIncomeIncentivesUcvsWc}
\end{eqnarray}

Finally, in order to reflect the different status of occupational classes, assume that:
\begin{equation}
\lambda_U > \lambda^W_U > \lambda_M.
\label{assumption:lambda_s}
\end{equation}

To summarize, the choice of individual $i$ hinges on the relationship between $\theta_i$ and $\lambda$s; in particular, if $\theta_i \in \left[0,\min\left( \lambda_M,\lambda^W_U\right) \right) = \left[0,\lambda_M\right)$ individual $i$ chooses to belong to Wc; if $\theta_i \in \left[\min\left( \lambda_M,\lambda^W_U\right),\max\left(\lambda^W_U, \lambda_U\right) \right) = \left[\lambda_M,\lambda_U\right)$ to belong to Mc; and, finally, if $\theta_i \in \left[\max\left(\lambda^W_U, \lambda_U\right),1 \right] = \left[\lambda_U,1\right]t$ to belong to Uc.

\subsection{Equality of Opportunity}

To introduce in the model equality of opportunity assume that the expected level of $\theta_i$ is decided by parental occupational class of individual $i$ ($POC_i$). In particular, assume that the probability distribution of $\theta_{i}$ conditioned to the fact that individual $i$'s parents belong to Wc is:
\begin{equation}
f(\theta_i| POC_i \in \text{Wc})\sim \mathcal{U}\left(0,\theta^{\max}\right),
\label{eq:uniformDitr1}
\end{equation}
where $\mathcal{U}\left(0,\theta^{\max}\right)$ means an uniformly distributed random variable in the range $\left[0,\theta^{\max}\right]$, with $\theta^{\max} \in \left(0,1\right]$. If her parents instead belong to Mc the probability distribution of $\theta_{i}$ is:
\begin{equation}
f(\theta_i| POC_i \in \text{Mc}) \sim \mathcal{U}\left(\theta_M^{\min},\theta_M^{\max}\right),
\label{eq:uniformDitr2}
\end{equation}
with $\theta_M^{\min} \in \left[0,1\right)$, $\theta_M^{\max} \in \left(0,1\right]$, and $ \theta_M^{\min} < \theta_M^{\max}$.
Finally, if her parents belong to Uc the probability distribution of $\theta_{i}$ is:
\begin{equation}
f(\theta_i| POC_i \in \text{Uc}) \sim \mathcal{U}\left(\theta^{\min},1\right),
\label{eq:uniformDitr3}
\end{equation}
with $\theta^{\min} \in \left[0,1\right)$.

An intuitive way to catch the meaning of parameters $\theta^{\min}$, $\theta_M^{\min}$, $\theta_M^{\max}$, and $\theta^{\max}$ is to assume that individual attributes $\theta_i$ are the product of \textit{genetic attributes} $g_i$ and \textit{cultural/social attributes} $ c \text{\scriptsize \&} s_i$, i.e.:
\[
\theta_i = g_i \times c \text{\scriptsize \&} s_i.
\]
Moreover, assume that $g_i$ are drawn from an uniform distribution in the range $[0,1]$ (i.e. no genetic class effect is present) and $ c \text{\scriptsize \&} s_i$ is a non random factor.
Since for children with parents in Wc $\theta_i$ must satisfy Eq. (\ref{eq:uniformDitr1}), then $\theta_i= g_i \theta^{\max}$; $\theta^{\max}<1$ therefore proxies for the cultural/social attributes of individual $i$ and \textit{reduces} the impact of genetic attributes on her $\theta_i$. To the opposite, children with parents in Uc must satisfy Eq. (\ref{eq:uniformDitr3}) and hence $\theta_i= \theta^{\min} + g_i \left( 1-\theta^{\min}\right)$; $\theta^{\min}>0$ proxies for cultural/social attributes of individual $i$ and \textit{puts a lower bound} on the value of $\theta_i$ independent of the genetic attributes of individual $i$. It is straightforward to conduct the same reasoning for individuals in Mc.

\subsection{Transition Probabilities between Occupational Classes}

\begin{figure}[htbp]
\begin{center}
\caption{\small{\bf The distribution of $\theta$s, the value of parameters $\lambda_M$ and $\lambda_U$, and the transition probabilities of children with parents in different occupational classes.}}
\vspace{0.3cm}
\includegraphics[width=1\textwidth]{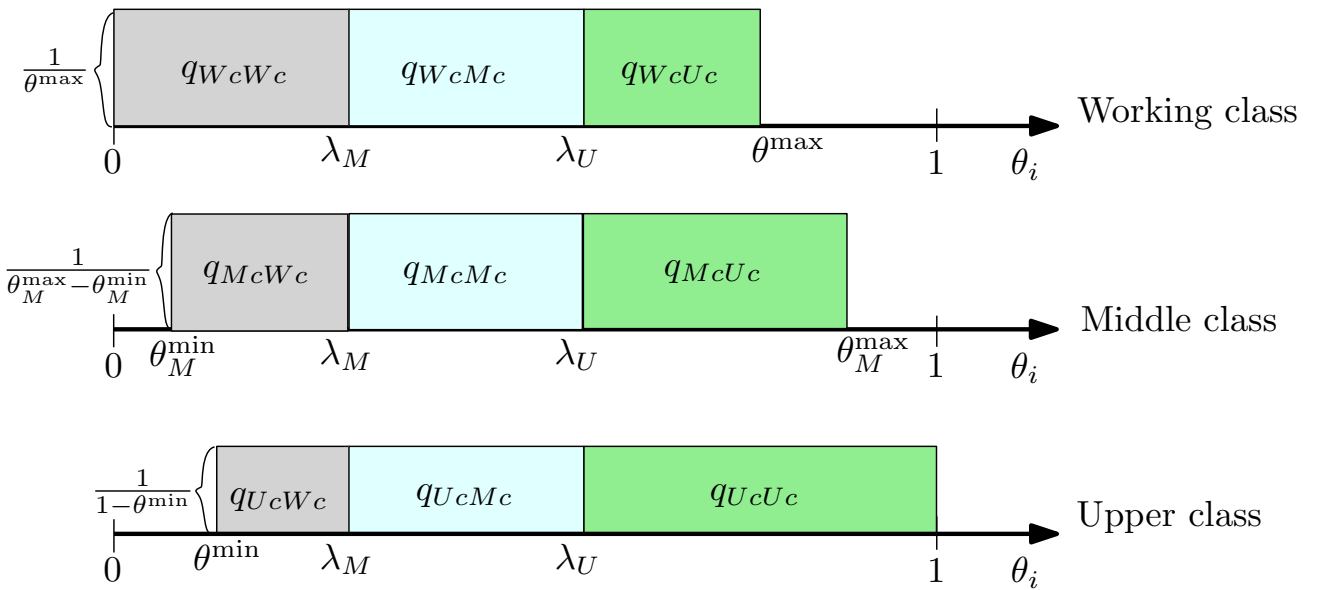}
\label{fig:OpportunitiesDistribution}
\end{center}
\end{figure}

The conditional distribution of $\theta$s and $\lambda$s are the base for the computation of the (transition) probability of a child with a parent in social class $i$ to belong to social class $j$, denoted by $q_{ij}$. For example, a child with a parent in Wc will remain in Wc if her $\theta_i$ is lower than $\lambda_M$; this implies that the mass of probability of not changing her social class is equal to $q_{WcWc}=\lambda_M/\theta^{\max}$ (see top panel in Figure \ref{fig:OpportunitiesDistribution}).

Figure \ref{fig:OpportunitiesDistribution} suggests the assumptions to be imposed on the value of parameters to have a well defined transition probabilities, i.e.:
\begin{eqnarray}
\label{assunzione:fullMobilityWc}
\theta^{\max} &\geq& \lambda_U; \\
\label{assunzione:fullMobilityMc}
\theta_M^{\max} &\geq& \lambda_U \geq \lambda_M  \geq \theta_M^{\min}; \text{ and} \\
\label{assunzione:fullMobilityUc}
\lambda_M &\geq& \theta^{\min}.
\end{eqnarray}
Under Assumptions (\ref{assunzione:fullMobilityWc})-(\ref{assunzione:fullMobilityUc}) the occupational mobility of a society ascribable to income incentives and opportunity is completely described by the \textit{transition matrix} $\mathbf{Q}$ reported in Table \ref{tab:MarkovMatrixQ}.\footnote{It is straightforward to find conditions under which $\mathbf{Q}$ is a monotone transition matrix (see \citealp{Dardanoni1995}).}
\begin{table}[h!]
\caption{\small{\bf Transition matrix with Working (W), Middle (M) and Upper (U) classes describing the true occupational mobility of society.}}
\begin{center}
$\mathbf{Q} \equiv  \bbordermatrix{ &  \text{\small{Wc (Children)}} & \text{\small{Mc (Children)}} & \text{\small{Uc (Children)}} \cr
	\text{\small{Wc (Parents)}} & \dfrac{\lambda_M}{\theta^{\max}} & \dfrac{\lambda_U-\lambda_M}{\theta^{\max}} & \dfrac{\theta^{\max}-\lambda_U}{\theta^{\max}} \cr
	\text{\small{Mc (Parents)}} & \dfrac{\lambda_M-\theta_M^{\min}}{\theta_M^{\max}-\theta_M^{\min}} & \dfrac{\lambda_U-\lambda_M}{\theta_M^{\max}-\theta_M^{\min}} & \dfrac{\theta_M^{\max}-\lambda_U}{\theta_M^{\max}-\theta_M^{\min}} \cr
	\text{\small{Uc (Parents)}} & \dfrac{\lambda_M-\theta^{\min}}{1-\theta^{\min}} & \dfrac{\lambda_U-\lambda_M}{1-\theta^{\min}} & \dfrac{1-\lambda_U}{1-\theta^{\min}} \cr
	}
$
\end{center}
\label{tab:MarkovMatrixQ}
\end{table}
By Figure \ref{fig:OpportunitiesDistribution} we can easily calculate all elements of $Q$. For example, the first element in the main diagonal of $\mathbf{Q}$ represents the probability of children with parents in Wc to belong to Wc, i.e.:
\begin{equation}
q_{WcWc} \equiv \mathbf{Pr}\left[\theta_i\leq\lambda_M| POC_i \in \text{Wc}\right]=\dfrac{\lambda_M}{\theta^{\max}},
\end{equation}
calculated on the base of Conditioned Probability Distribution (\ref{eq:uniformDitr1}) of $\theta_i$ and the children's incentives to belong to Wc given by Condition (\ref{eq:ConditionIncomeIncentives}). 

\subsubsection{Perfect Mobile Versus Perfect Immobile Societies}

Matrix $\mathbf{Q}$ in Table \ref{tab:MarkovMatrixQ} allows to identify two polar cases of societies, already discussed in \citet{Prais1955}, i.e. the Perfect Mobile Society (PMS) and the Perfect Immobile Society (PIS).
In PMS the probability to belong to a occupational class is independent of the class of own parents. This happens when all conditioned distributions of $\theta_i$ are the same and not biased, i.e. for $\theta^{\max}=\theta^{\max}_M=1$ and $\theta^{\min} = \theta^{\min}_M =0$, and hence:
\begin{equation}
\mathbf{Q_{PMS}} =
\begin{bmatrix}
    \lambda_M  & \lambda_U-\lambda_M & 1-\lambda_U\\
	\lambda_M  & \lambda_U-\lambda_M & 1-\lambda_U\\
	\lambda_M  & \lambda_U-\lambda_M & 1-\lambda_U\\
  \end{bmatrix},
  \label{eq:PMS}
\end{equation}
Moreover, the couple of parameters $ \left(\lambda_M,\lambda_U\right)$ defines alternative PMSs, which can be ranked in terms of their social welfare; in particular, for $\lambda_M=1/3$ and $\lambda_U=2/3$:
\begin{equation}
\mathbf{Q_{PMS}^{S}}=\left[
\begin{array}{ccc}
    1/3 & 1/3 & 1/3 \\
    1/3 & 1/3 & 1/3 \\
    1/3 & 1/3 & 1/3 \\
  \end{array}
\right],
\label{eq:EquityMatrix}
\end{equation}
which represents a society with a symmetric downward and upward occupational mobility; on the contrary, for $\lambda_M=\lambda_U=0$:
\begin{equation}
\mathbf{Q_{PMS}^{U}}=\left[
\begin{array}{ccc}
     0  & 0 & 1\\
     0  & 0 & 1\\
     0  & 0 & 1 \\
  \end{array}
\right].
\label{eq:PerfectMobilityMatrixUpper}
\end{equation}
a society with just an upward mobility toward the most prestigious Upper class. In terms of social welfare, therefore, $\mathbf{Q_{PMS}^{U}} \succ \mathbf{Q_{PMS}^{S}}$.

The opposite case is the Perfect Immobile Society (PIS), where no movements between classes take place, which happens for $\theta^{\max} \leq \lambda_M$, $\theta_M^{\min} \geq \lambda_M$, $\theta_M^{\max} \leq \lambda_U$,  $\theta^{\min} \geq \lambda_U$, i.e.:
\begin{equation}
  \mathbf{Q_{PIS}} =\begin{bmatrix}
      1  & 0 & 0\\
      0  & 1 & 0\\
      0  & 0 & 1\\
    \end{bmatrix}.
    \label{eq:PIS}
\end{equation}

\subsection{Occupational Shifts \label{sec:DecompositionMobility}}

In a pioneering contribution \citet{Prais1955} argues that the observed occupational mobility is the outcome of two distinct phenomena: i) the choices of individuals, which in the model is denoted as \textit{true mobility}; and ii) the \textit{occupational shifts}, that is the changes in the occupational structure caused both by the mutations in the production side of economy and by the differences in the fertility rates among occupational classes, which in our model is denoted as \textit{structural mobility}. In particular, \citet{Prais1955} assumes that the \textit{observed} transition matrix $\mathbf{P}$ can be expressed as the result of the product of two transition matrices: $\mathbf{Q}$, representing true mobility, and $\mathbf{R}$, representing structural mobility, i.e.:
\begin{equation}
\mathbf{P} = \mathbf{Q} \mathbf{R}.
\label{eq:disantenglingTrueOccMobilityFromOccShifts}
\end{equation}

In order to clarify the meaning of $\mathbf{R}$, denote by $s_{t}$ the vector of the shares of individuals in each occupational class at period $t$; in the case \textit{unconstrained occupational mobility} the dynamics of these shares is given by:
\begin{equation}
s_{t+1,UN}=\mathbf{Q^\top}s_{t},
\label{eq:dynamicsIndividualsShareOccupationalClasses}
\end{equation}
where $s_{t+1,UN}$ is the vector of shares in absence of constraints on the production side and no difference in fertility rates among occupational classes.
The difference between $s_{t+1,UN}$ and $s_{t+1}$ can be expressed by a transition matrix $\mathbf{R}$, which proxies the importance of occupational shifts in the observed occupational mobility, i.e.:
\begin{equation}
s_{t+1}=\mathbf{R^\top}s_{t+1,UN},
\label{eq:decompositionMobility1}
\end{equation}
which, substituted from Eq. (\ref{eq:dynamicsIndividualsShareOccupationalClasses}) for $s_{t+1,UN}$, yields:
\begin{equation}
s_{t+1} = \mathbf{R^\top}\mathbf{Q^\top}s_{t}= \mathbf{P^\top}s_{t},
\label{eq:decompositionMobility2}
\end{equation}
where we use Eq. (\ref{eq:disantenglingTrueOccMobilityFromOccShifts}).

In the case of no occupational shifts $\mathbf{R}=\mathbf{I}$, where $\mathbf{I}$ is the identity matrix; this implies that the individual choices do not face any constraint from the production side and by differential fertility rates among occupational classes. Differently, if $s_{t+1,Wc}>s_{t,Wc}$ and $s_{t+1,Mc}<s_{t,Mc}$ and $s_{t+1,Uc}<s_{t,Uc}$, i.e. occupational shifts are in favour of the share of Wc, some children who would like to belong to Mc and Cc are constrained in their choice to Wc (increasing occupational mobility) because production and/or fertility are favoring occupations in Wc at the expense of occupations in Mc and Cc. In a growing economy with structural change in favor of information and technological sector, we expect the opposite dynamics.

In order to estimate $\mathbf{Q}$ we must know $\mathbf{R}$ and $\mathbf{P}$, but only $\mathbf{P}$ can be directly estimated from data. The estimate of $\mathbf{R}$ is possible only imposing some identifying assumptions. In particular, \citet{Prais1955} proposes a criterion of \textit{individually minimum occupational mobility} to identify $\mathbf{R}$, and implements such criterion by an algorithm starting from the bottom class of parents (our Wc), and \textit{sequentially} arriving to the top one (our Uc), allocating children in each class minimizing the change between their occupational class and the one of their parents. We instead estimate $\mathbf{R}$ under the criterion of \textit{globally minimum occupational mobility}, where occupational mobility is measured by the opposite of the trace of $\mathbf{R}$, subject to the observed occupational shifts, i.e.:
\begin{equation}
\begin{aligned}
& \underset{\mathbf{R}}{\min} - \tr\left(\mathbf{R}\right) &  \\
&   \text{s. t.} \begin{cases}s_{t+1}= \textbf{R}^{\top}s_t\\
 \sum_{j=1}^k r_{ij}=1 & \forall i=1,\ldots,k\\
 r_{ij}\geq0 &  \; \forall i,j. \end{cases}
 \end{aligned}
 \label{eq:MaximizeProblem}
\end{equation}
We prefer our criterion with respect to the one proposed by \citet{Prais1955} because it is more coherent with our measure of occupational mobility based on the trace of transition matrix discussed in Section \ref{sec:MeasureOccupationalMobility}.\footnote{In the procedure for the calculation of $r_{ij}$ we need that $q_{ij} \geq 0 \; \forall i,j$. The following procedure allows to satisfy the latter constrains: 1) calculate $\mathbf{R}$ on the base of Algorithm (\ref{eq:MaximizeProblem}); 2) calculate $\mathbf{Q}'=\mathbf{P}\mathbf{R}^{-1}$; 3) setting $q''_{ij}=q'_{ij}$ if $q'_{ij}\geq 0$ or $q''_{ij}=0$ if $q'_{ij} < 0$; 3) row normalizing the resulting matrix, i.e. $q'''_{ij} = q''_{ij}/\sum_{j=1}^{k}q''_{ij}$ (this amounts to proportionally decrease the excessive mass of non null elements of $\mathbf{Q}'$); 4) calculate $\mathbf{R}'=\mathbf{Q}'''^{-1}\mathbf{P}$; 5) calculate $\mathbf{R}''$ by row-normalizing $\mathbf{R}'$ in order to satisfy $ \sum_{j=1}^k r_{ij}=1 \; \forall i=1,\ldots,k$; and, finally, 6) calculate $\mathbf{Q}''''=\mathbf{P}\mathbf{R}''^{-1}$.}
Finally,  we observe that algorithm for the estimate of $\mathbf{R}$ tends to positively bias the estimate of true mobility.

\subsection{Indexes of Occupational Mobility \label{sec:MeasureOccupationalMobility}}

In this Section we illustrate how the different types of mobility discussed so far, true versus structural mobility, and the decomposition between income incentive and equality of opportunity, can be measured by synthetic indexes.
The framework of these indexes is based on Eq. (\ref {eq:disantenglingTrueOccMobilityFromOccShifts}), which expresses the observed mobility as:
\begin{equation}
I_{OBS}=I_{TRUE} \times I_{OS},
\label{eq:decompositionObservedMobility}
\end{equation}
i.e. the index of observed mobility $I_{OBS}$ is expressed as the product of an index of true mobility $I_{TRUE}$ and an index of occupational shifts $I_{OS}$. In turn, $I_{TRUE}$ is the difference between an index of the equality of opportunity $I_{OPP}$ and an index of the lack of incentives to exploit these opportunities $I_{LOI}$, i.e.
\begin{equation}
I_{TRUE} = I_{OPP}-I_{LOI}.
\label{eq:decompositionTrueMobility}
\end{equation}
Eq. (\ref{eq:decompositionTrueMobility}) represents a novelty with respect the existent literature (see, e.g., \citealp{ChecchiPeragine2010} and \citealp{BrunoriEtAl2013}), where equality of opportunity is generally related to ex-post inequality and not to the choice set of individuals. In particular, \citet{ChecchiPeragine2010} propose an index of inequality as the sum of an index of opportunity and an index of effort, where their index of opportunity is \textit{inversely} related to $I_{OPP}$.

\subsubsection{Observed, True, and Structural Mobility}

Inspired by \citet{Shorrocks1978}, as index of observed occupational mobility we take the complement to $3$ of the trace of transition matrix among different occupations, on the intuition that a lower value of trace corresponds to a higher occupational mobility, i.e. to a higher probability for a children to move out from the occupational class of parents.\footnote{The original Shorrocks index is divided by 2 and not by 3, being defined as $\left(k-\tr\left(Q\right)\right)/(k-1)$.} In particular,
\begin{equation}
I_{OBS} \equiv 1 - \dfrac{\tr\left(\mathbf{P}\right)}{3},
\label{eq:ShorrocksIndexObservedMobility}
\end{equation}
with $I_{OBS} \in \left[0,1\right]$, measures the intensity of the \textit{observed} occupational mobility (0 minimum, and 1 maximum observed occupational mobility); in the same fashion, we propose as index of true occupational mobility:
\begin{equation}
I_{TRUE} \equiv 1 -\dfrac{\tr\left(\mathbf{Q}\right)}{3},
\label{eq:ShorrocksIndexTrueOccupatinalMobility}
\end{equation}
with $I_{TRUE} \in [0,1]$. Finally, we take as index of occupational shifts:\footnote{We observe that if $I_S$ were calculated as $1 -\tr\left(\mathbf{R}\right)/3$ then Eq. (\ref{eq:decompositionObservedMobility}) should be not respected given that $\tr(A/B) \neq tr(A)/tr(B)$.}
\begin{equation}
I_{OS} \equiv \dfrac{I_{OBS}}{I_{TRUE}}
\label{eq:ShorrocksIndexStructuralMobility}
\end{equation}
From Table \ref{tab:MarkovMatrixQ} we can calculate $I_{TRUE}$:
\begin{equation}
I_{TRUE} = 1 - \left(\dfrac{1}{3}\right)\left[\dfrac{\lambda_M}{\theta^{\max}} + \dfrac{\lambda_U-\lambda_M}{\theta_M^{\max}-\theta_M^{\min}} + \dfrac{1-\lambda_U}{1-\theta^{\min}}\right].
\label{eq:ShorrocksIndexTrueOccupatinalMobilityModelsParameters}
\end{equation}
As expected $I_{TRUE}$ is a positive function of $\theta^{\max}$ and $\theta_M^{\max}$ and a negative function of $\theta_M^{\min}$ and $\theta^{\min}$; the impact of $\lambda_M$ and $\lambda_U$ instead is ambiguous: an increase in $\lambda_M$ ($\lambda_U$) causes both an increase in the probability to remain in the same class for children with parents in Wc (Mc) and, at the same time, a decrease in the same probability for children with parent in Mc (Uc). 

\subsubsection{Equality of Opportunity and Lack of Incentives}

\begin{figure}[h]
\begin{center}
\caption{\small{\bf The decomposition of true mobility's index $I_{TRUE}$.}}
	\vspace{0.3cm}
\includegraphics[width=1\textwidth]{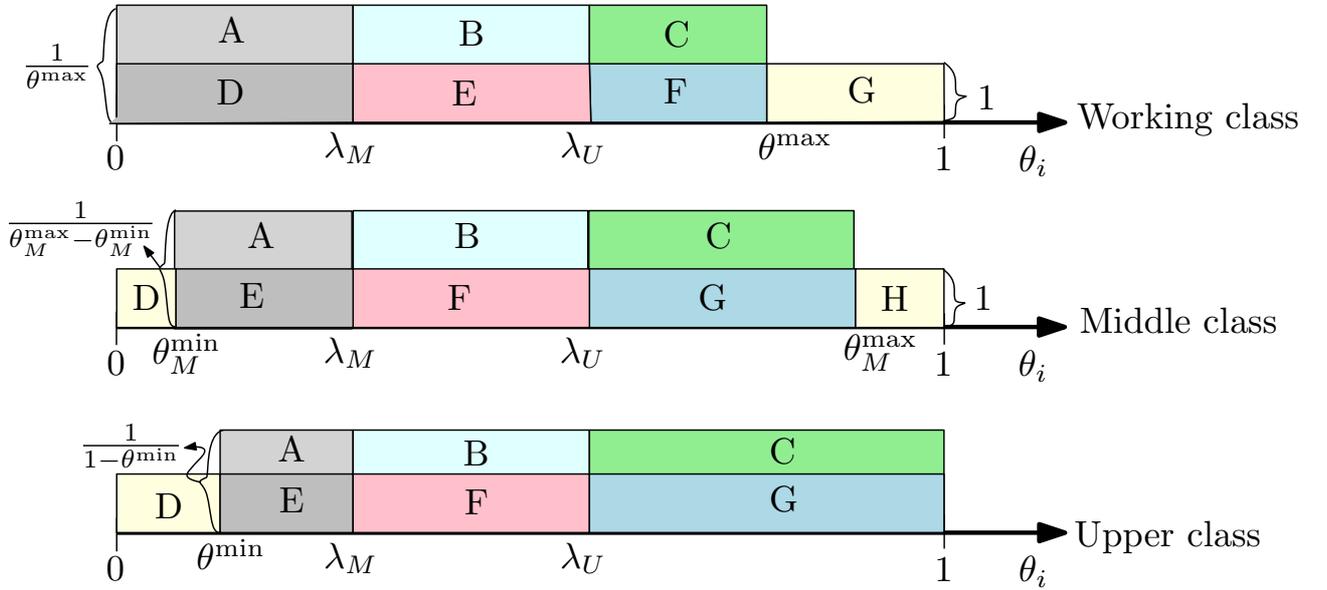}
\label{fig:decompostionTrueMobilityWcMcUc}
\end{center}
\end{figure}

The calculation of equality of opportunity's index requires an additional reasoning with the help of Figure \ref{fig:decompostionTrueMobilityWcMcUc}. 

In particular, for children with parents in Wc, Area G proxies for the lower probability of moving from the class of their parents due to the bias in the distribution of $\theta_i$. This suggests to measure the contribution of Wc to overall index of the equality of opportunity $I_{OPP}$ by 1-G (G goes to zero, i.e. max equality of opportunities, as $\theta^{\max}$ goes to 1). Moreover, since the contribution of Wc to true mobility index $I_{TRUE}$ is given by Areas B+C+E+F=1-A-D, then the contribution of Wc to the overall index of the lack of incentive to exploit these opportunities $I_{LOI}$ is given by Areas D+A-G. 

By making the same reasoning for Mc and Uc, averaging the contributions of each class, and calculating the areas on the base of our theoretical model we get that:
\begin{equation}
I_{OPP} = \dfrac{1+ \theta^{\max} + \theta_M^{\max} - \theta^{\min} - \theta_M^{\min}}{3}
\label{eq:indexEqualityOpportunity}
\end{equation}
and
\begin{equation}
I_{LOI} = \left(\dfrac{1}{3}\right)\left[ 1 + \theta^{\max} + \theta_M^{\max} - \theta^{\min} - \theta_M^{\min} + \dfrac{\lambda_M}{\theta^{\max}} + \dfrac{\lambda_U-\lambda_M}{\theta_M^{\max}-\theta_M^{\min}} + \dfrac{1-\lambda_U}{1-\theta^{\min}}\right]  - 1.
\label{eq:indexLackOfIncentive}
\end{equation}
According to Eq. (\ref{eq:indexEqualityOpportunity}) $I_{OPP}$ reaches the maximum value of 1 for $\theta^{\max} = \theta_M^{\max}=1$ and $\theta^{\min} = \theta_M^{\min} = 0$, i.e. where no bias are present in the distribution of $\theta_i$. In particular, $I_{OPP}=1$ for each true mobility matrix corresponding to a Perfect Mobile Society, i.e. $Q=Q_{PMS}$ (see Matrix \ref{eq:PMS}).

\subsection{The Identification of Mobility and Equality of Opportunity\label{sec:EstimateParametersModel}}

The micro-foundation of transition matrix allows to circumvent the key limitation in the identification of mobility and equality of opportunity generally due to the lack of data (\citealp{RoemerTrannoy2016}).
To appreciate the issue consider that in our model a higher persistence in Wc class, i.e. a higher $\lambda_M/\theta^{\max}$ (see Table \ref{tab:MarkovMatrixQ}), can be the outcome of two possible dynamics: i) a decline in income incentive to move to higher classes, i.e. a higher $\lambda_M$; and ii) a decline in the equality of opportunity, i.e. a lower $\theta^{\max}$. While the individual element is not sufficient to determine the true cause of the higher persistence, identification is possible considering all elements of $\mathbf{Q}$ since the latter has six independent elements, exactly the same number of model's parameters.\footnote{In $\mathbf{Q}$ the constraint that the sum of all elements on each row must be equal to 1 makes independent only six elements.} In particular:
\begin{eqnarray}
\begin{cases}
\lambda_M  =  \dfrac{ q_{32}/q_{33} }{ \left(1+q_{12}/q_{11}\right)\left(1+q_{32}/q_{33}\right)-1}; \\
\lambda_U = \lambda_M \left(1+q_{12}/q_{11}\right); \\ 
\theta^{\max} = \dfrac{\lambda_M}{q_{11}}; \\ 
\theta^{\min} = 1- \dfrac{1-\lambda_U}{q_{33}}; \\
\theta^{\min}_M = \lambda_M - \left(\lambda_U-\lambda_M\right)\left(\dfrac{q_{21}}{q_{22}}\right); \text{ and} \\
\theta^{\max}_M = \theta^{\min}_M + \dfrac{\lambda_U-\lambda_M}{q_{22}}.
\label{eq:identificationParameters}
\end{cases}
\end{eqnarray}
In other words, the system of equations directly derived by the definition of the elements of transition matrix is sufficient to estimate the equality of opportunity and income incentives with an information set corresponding to the case of \textit{poor dataset} in \cite{RoemerTrannoy2016}.

\section{An Estimate of Occupational Mobility in Italy \label{sec:EmpiricalAnalysis}}

In this Section we estimate our theoretical model for a sample of Italian heads of household born in the period 1940-1976. Section \ref{sec:Dataset} describes the dataset in more details, Section \ref{sec:definitionOccupationalClasses} provides the definition of occupational classes, and Section \ref{sec:EstimateItalianMobility} contains the estimates and a discussion of the main findings.\footnote{Dataset and codes are available at Davide Fiaschi's website \url{https://people.unipi.it/davide_fiaschi/ricerca/}.}

\subsection{The Dataset \label{sec:Dataset}}

\begin{figure}[ht]
\begin{center}
\caption{\small{\bf A description of socioeconomic conditions of Italy when children within each cohort took their choices on education and occupation (assumed be at about at 15 years old).}}
	\vspace{0.3cm}
\includegraphics[width=0.8\textwidth]{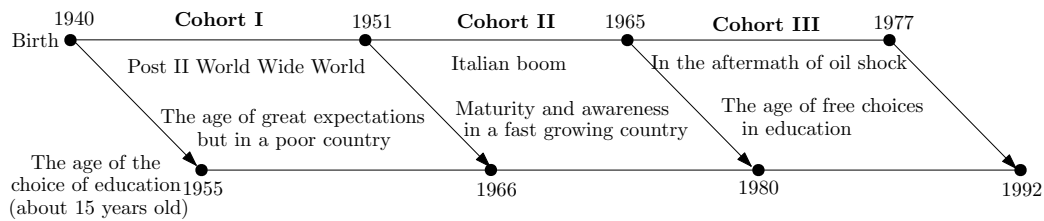}
\label{fig:interpretazioneDiverseCorti}
\end{center}
\end{figure}

The dataset is build from a nationally representative household survey carried on by the Bank of Italy called the ``\emph{Survey on Household Income and Wealth}'' (SHIW).
We pool the nine waves conducted in the period 1995-2012, selecting all heads of household aged from 35 up to 65 (i.e born between 1940 and 1977). In all these waves the heads of household are asked to recall some characteristics of their parents, among which the year of birth and the occupational status, indicatively referred to the same current age of the respondent, which allows to control for the \textit{life cycle component} of occupational mobility. We measure occupational mobility comparing occupational status of children with those of their \textit{fathers} (see \citealp{Checchi1997} and \citealp{Piraino2007}). We removed the heads of household not giving informations on their fathers (about 15\% of sample) and the repeated observations due to longitudinal component (panel) of the waves (about 30\% of households persists from a wave to the next one). Overall, the sample consists of 14,037 observations. In order to study the dynamics of occupational mobility over time we then partition the sample into three cohorts on the base of the year of birth of the heads of household. The first cohort includes 4,166 observations of the heads of household born in the period 1940-1951 (Cohort I); the second  7,696 observations of those born between 1952 and 1965 (Cohort II); and the third 2,175 observations of those born between $1966$ and $1977$ (Cohort III). The ratio of this division into three cohorts is exposed in Figure \ref{fig:interpretazioneDiverseCorti}.

Cohort I includes children who took their choices on education and occupation (assumed to be at about 15 years old) in the aftermath of the II World War (between 1955 and 1966) in a poor but fast growing economy, with a strong structural change in favor of manufacturing sectors. This environment encouraged occupational mobility, especially thanks to the production side and to favorable economic prospects, while credit constraints and cultural/social were the main braking factors. Cohort II collects children taking their choices in the period (between 1966 and 1980) with the highest growth rate of income in the last century, with a increasing maturity and awareness on educational and working decisions. All these factors are conductive to an higher occupational mobility toward Mc and Uc. Finally, Cohort III gathers children who took their choices more freely, but also in a period (post 1992) where occupational structure was stable and expectations about economy were gloomy. Occupational mobility toward Mc and Uc should have been negatively affected.
 
\subsection{The Definition of Occupational Classes \label{sec:definitionOccupationalClasses}}

The definition of which occupations are included in Wc, Mc, and Uc is based on \citet{GoldthorpeHope1974}, who rank occupational classes according to their social prestige (see also Cap. 12 in \citealp{GiddensSutton2013}). In particular, \textit{Hope-Goldthorpe scale} consists of eight classes (blue-collars, clericals, teachers, managers, member of profession, entrepreneurs and self-employment workers), and mainly reflects the average income paid by each occupation, but also other criteria more related to social status. We consider the socio-economic classes of blue-collars as Wc, clericals, teachers, self-employment workers as Mc, and managers, member of profession, and entrepreneurs as Uc (see Cap. 12 in \citealp{GiddensSutton2013} for a detailed discussion).\footnote{In SHIW the occupational class of children is coded by label APQUAL. Children whose answers are: i) 1 or 20 (blue-collar and atypical workers respectively) are assigned to Wc; ii) 2, 3, 4, or 8 (clerk, teacher, office worker, junior manager, small employer, own account worker respectively) are assigned to Wc; and, iii) 5, 6, 7, or 10 (senior manager, official, school head, professor, magistrate, member of professions, entrepreneur, administrator respectively). Unemployed children are assigned to the class of their last job if available. The occupational class of fathers is coded by label CONPCF; we use the same reclassification of children.} 
In terms of ISCO classification, Wc corresponds about to ISCO major groups 6-9, Mc 3-5, and Uc 1-2, which is the same partition used by \citet{FranziniEtAl2013}.

 \subsection{The Estimate of Occupational Mobility and its Determinants \label{sec:EstimateItalianMobility}}

Table \ref{tab:Transitionprobabilities_estimate} reports the estimate of $\mathbf{P}$, $\mathbf{R}$ and $\mathbf{Q}$ for each cohort.\footnote{In particular, we first estimate $\mathbf{P}$; then, on the base of the observed occupational shifts and the solution of Problem (\ref{eq:MaximizeProblem}) in Section \ref{sec:DecompositionMobility} (amended to prevent negative elements in $\mathbf{Q}$), $\mathbf{R}$. Finally, by Eq. (\ref{eq:decompositionMobility2}) we get the estimate of $\mathbf{Q}$.} The observed occupational mobility appears generally low for all three classes with respect to to other developed countries (see, e.g., \Citealp{Jonsson2004} for Sweden, \Citealp{MullerPollak2004} for Germany, \Citealp{GanzeboomLuijkx2004} for Netherlands, and \Citealp{ChecchiEtAl1999} for US); between 52\% and 60\% of children  of Cohorts II and III with fathers in Wc and Mc remain in the same parent's class (see the elements along the main diagonal of estimated $\mathbf{P}$). This evidence further supports the results of \citet{SchizzerottoCobalti1994}, \citet{Pisati2000}, \citet{DiPietroUrwin2003}, and of more recent contribution by \citet{RaitanoVona2015}, which suggests high persistence in the lowest occupational classes in Italy. In particular, we find a still worse picture of the phenomenon in the youngest cohort, with an estimated probability to persist in Wc equal to 60\% against one of 51\% found in \citet{Pisati2000}. Moreover, the high and steady persistence in Mc, about 60\% in all three cohorts, is the result of an increased downward mobility (from 22\% to 28\% from Mc to Wc) and a corresponding decreased upward mobility (from 18\% to 12\% from Mc to Uc). A decreased upward mobility is experimented also by children with fathers in Wc (from 42\% to 35\% from Wc to Mc and from 9\% to 5\% from Wc to Uc), while an increased downward mobility by children with fathers in Uc (from 35\% to 55\% from Uc to Mc and from 8\% to 13\% from Uc to Wc). In short, the overall picture on Italian occupational mobility is disappointing: a high persistence in the low occupational classes Wc and Uc, associated to an increasing downward mobility and to a decreasing upward mobility. 

As occupational shifts, the estimate of $\mathbf{R}$ in Table \ref{tab:Transitionprobabilities_estimate} highlights the important, but declining, role of production side in favoring the escape from Wc in favor of Mc and Uc, and from Mc in favor of Uc. More precisely, the share of children with fathers in Wc who moves from Wc because of the change in the occupational structure of Italian economy goes from 26\% of Cohort I to 16\% of Cohort III, while the share of children with fathers in Mc who moves from Uc goes from 5\% of Cohort I to 0 of Cohort III. The comparison between occupational structure of children and fathers for the three cohorts reported in Table \ref{tab:occupationalStructure} further confirms that occupational shifts less and less favor upward mobility over time (for example, in Cohort I 51\% of fathers belongs to Wc versus 35\% of children, while in Cohort III 52\% versus 44\% of children).

The true occupational mobility as estimated by $\mathbf{Q}$ in Table \ref{tab:Transitionprobabilities_estimate} appears lower than observed mobility: the persistence in Wc is much higher, peaking at 72\% for Cohort III, and it is almost the same in Mc and Uc (56\% and 31\% in Cohort III respectively). Moreover, upward mobility from Wc is decreasing from Cohort I to Cohort III, while is increasing the downward mobility from Uc, peaking at 53\% to Mc and 15\% to Wc for Cohorts III. Overall, we find the same picture of observed mobility: high persistence in Wc and Uc associated to an increasing downward mobility.

\begin{table}[htbp]
\centering
\caption{\small{\bf{The estimate of matrices $\mathbf{P}$ (observed mobility), $\mathbf{R}$ (structural mobility), and $\mathbf{Q}$ (true mobility) for children in Cohorts I, II and III.}}}
\begin{threeparttable}
\small{
\begin{tabular}{l|rrrH|rrrH|rrrH}
  \hline 
  \hline
  \multicolumn{5}{c}{\hspace{2cm} Matrix \textbf{P}} & \multicolumn{4}{c}{Matrix \textbf{R}} & \multicolumn{4}{c}{Matrix \textbf{Q}} \\
    \hline
Cohort I & Wc & Mc & Uc & Matrix P & Wc & Mc & Uc & Matrix R & Wc & Mc & Uc & Matrix Q \\ 
  \hline 
  Wc & 0.49 & 0.42 & 0.09 & 1894 & 0.74 & 0.15 & 0.10 & 1894 & 0.67 & 0.33 & 0 & 1894 \\ 
  Mc & 0.22 & 0.60 & 0.18 & 1639 & 0 & 0.95 & 0.05 & 1639 & 0.29 & 0.59 & 0.12 & 1639 \\ 
  Uc & 0.08 & 0.35 & 0.57 & 212 & 0.01 & 0.01 & 0.99 & 212 & 0.10 & 0.35 & 0.55 & 212 \\ 
  Obs. & 1307 & 1847 & 591 & 3745 & 1307 & 1847 & 591 & 3745 & 1307 & 1847 & 591 & 3745 \\
  \hline
  Cohort II & Wc & Mc & Uc & Matrix P & Wc & Mc & Uc & Matrix R & Wc & Mc & Uc & Matrix Q \\ 
  \hline 
  Wc & 0.52 & 0.41 & 0.07 & 4448 & 0.72 & 0.19 & 0.09 & 4448 & 0.72 & 0.28 & 0 & 4448 \\ 
  Mc & 0.22 & 0.61 & 0.17 & 3356 & 0 & 0.97 & 0.03 & 3356 & 0.31 & 0.56 & 0.13 & 3356 \\ 
  Uc & 0.10 & 0.49 & 0.41 & 521 & 0.01 & 0.01 & 0.98 & 521 & 0.14 & 0.47 & 0.39 & 521 \\ 
  Obs. & 3104 & 4120 & 1101 & 8325 & 3104 & 4120 & 1101 & 8325 & 3104 & 4120 & 1101 & 8325 \\ 
  \hline
  Cohort III & Wc & Mc & Uc & Matrix P & Wc & Mc & Uc & Matrix R & Wc & Mc & Uc & Matrix Q \\ 
  \hline
  Wc & 0.60 & 0.35 & 0.05 & 1736 & 0.84 & 0.12 & 0.04 & 1736 & 0.72 & 0.26 & 0.02 & 1736 \\ 
  Mc & 0.28 & 0.60 & 0.12 & 1327 & 0 & 1 & 0 & 1327 & 0.33 & 0.56 & 0.11 & 1327 \\ 
  Uc & 0.13 & 0.55 & 0.32 & 256 & 0 & 0 & 1 & 256 & 0.15 & 0.53 & 0.31 & 256 \\ 
  Obs. & 1451 & 1542 & 326 & 3319 & 1451 & 1542 & 326 & 3319 & 1451 & 1542 & 326 & 3319 \\ 
   \hline
   \hline
\end{tabular}
}
\begin{tablenotes}
	\item[] \footnotesize{\textit{Note:} Cohort I is made up of children whose fathers are born in $1940-1951$, Cohort II of children whose fathers are born in $1952-1965$, and Cohort III of children whose fathers are born in $1966-1977$.}
	\item[] \footnotesize{\textit{Source:} Our calculations based on Survey on Household Income and Wealth (Bank of Italy).}
\end{tablenotes}
\end{threeparttable}
\label{tab:Transitionprobabilities_estimate}
\end{table}

\begin{table}[htbp]
	\centering
	\caption{\small {\bf The occupational structure of fathers and children for Cohort I, II and III.}}
	\begin{threeparttable}
\begin{tabular}{l|rrr|rrr|rrr}
  \hline
  \hline
  \multicolumn{4}{c}{\hspace{2.5cm}   Cohort I} & \multicolumn{3}{c}{Cohort II} & \multicolumn{3}{c}{Cohort III} \\
  \hline
  & Wc & Mc & Uc & Wc & Mc & Uc & Wc & Mc & Uc \\ 
  \hline
  Fathers & 0.51 & 0.44 & 0.06 & 0.53 & 0.40 & 0.06 & 0.52 & 0.40 & 0.08 \\ 
  Children & 0.35 & 0.49 & 0.16 & 0.37 & 0.49 & 0.13 & 0.44 & 0.46 & 0.10 \\ 
   \hline
   \hline
\end{tabular}
\begin{tablenotes}
	\item[] \footnotesize{\textit{Source:} Our calculations based on Survey on Household Income and Wealth (Bank of Italy).}
\end{tablenotes}
\end{threeparttable}
\label{tab:occupationalStructure}
\end{table}


The estimates of the indexes of occupational mobility reported in Table \ref{tab:estimateIndexOccupationalMobility} calculated by Eq. (\ref{eq:ShorrocksIndexObservedMobility})-(\ref{eq:ShorrocksIndexStructuralMobility}) depict a growing occupational mobility, raised from 0.45 for Cohort I to 0.49 for Cohort III, which we know from the above discussion to be largely ascribable to an increasing downward mobility, especially in Uc, a declining structural mobility (from 1.11 for Cohort I to 1.05 for Cohort III), caused by the fading structural change in favor of middle and upper classes; and, finally, a rising true mobility climbed by 0.4 for Cohort I to 0.46 for Cohort III, again largely caused by the increased downward mobility in Uc.\footnote{All the differences are statistical significant at usual 5\% significance level.} 

Summarizing, observed occupational mobility is increased from Cohort I to Cohort III, but unfortunately for ``bed'' reasons, i.e. an increased downward mobility, especially from Uc. Occupational shifts, which favored the escape from Wc, faded over time making the true occupational mobility lower, but, again, for ``wrong'' reasons, as a result of higher persistence in Wc and a higher downward mobility both Mc and Uc. Our results are in line with the evidences discussed by other empirical studies. \citet{Goldberger2016}, analysing social mobility in Britain over the period from the end of the Second World War down to the early 21st century, highlights that the total occupational mobility has remain stable but, distinguishing between upward and downward mobility, the former is decreased, whereas the latter is increased. Moreover, for U.S. and four Europeans countries (Germany, France, Sweden and Netherlands) \citet{Breen2019} finds a pattern of increasing downward mobility and of decreasing upward mobility for those individuals born between 1960 and 1980.

\begin{table}[htbp]
	\centering
	\caption{\small {\bf Estimated indexes of observed, structural and true occupational mobility for Cohort I, II and III.}}
	\begin{threeparttable}
		\begin{tabular}{l|ccc}
			\hline
			\hline
			& $I_{OBS}$ & $I_{OS}$ & $I_{TRUE}$ \\ 
			\hline
			Cohort I & 0.45 & 1.11 & 0.40 \\[-1.5ex]
		
			& (0.012) & (0.049) & (0.013) \\[-1.0ex]
			Cohort II & 0.49 & 1.10 & 0.44 \\[-1.5ex]
		
			& (0.008) & (0.03) & (0.009) \\[-1.0ex]
			Cohort III & 0.49 & 1.05 & 0.47 \\[-1.5ex]
	
			& (0.014) & (0.045) & (0.015) \\
			\hline
			\hline
		\end{tabular}
		\begin{tablenotes}
			\item [] \footnotesize{\textit{Notes:} Standard errors of estimates, reported between brackets, are calculated by 1000 bootstraps on the observed occupational transitions used to estimate $\mathbf{P}$.}
			\item[] \footnotesize{\textit{Source:} Our calculations based on Survey on Household Income and Wealth (Bank of Italy).}
		\end{tablenotes}
	\end{threeparttable}
	\label{tab:estimateIndexOccupationalMobility}
\end{table}

\subsubsection{Disentangling the Effects of Equality of Opportunities from the Lack of Incentives on Mobility \label{sec:disentangling}}

The estimates of model's parameters in Table \ref{tab:estimatedParameters} made by Eqq. (\ref{eq:identificationParameters}) suggest that income incentives and (in)equality of opportunities played an important role for the mobility of all three cohorts, being $\theta^{\max}$ and $\theta^{\max}_M$ largely under 1, while $\theta^{\min}$ and $\theta^{\min}_M$ well above 0. \citet{Checchi2003} finds a similar crucial role of family background for educational attainments.

\begin{table}[htbp]
\centering
\caption{\small{\bf Estimate of the model's parameters and of the indexes of equality of opportunities $I_{OPP}$ and lack of incentives $I_{LOI}$.}}
\begin{threeparttable}
\begin{tabular}{l|cccccc|ccc}
  \hline
   \hline
  & $\lambda_M$ & $\lambda_U$ & $\theta^{\max}$ & $\theta^{\min}$ & $\theta^{\max}_M$ & $\theta^{\min}_M$  & $I_{TRUE}$ & $I_{OPP}$ & $I_{LOI}$ \\ 
  \hline
Cohort I & 0.44 & 0.66 & 0.66 & 0.38 & 0.71 & 0.33 & 0.40 & 0.55 & 0.15 \\[-1.5ex]

		& (0.027) & (0.036) & (0.036) & (0.035) & (0.038) & (0.025) & (0.014) & (0.013) & (0.003) \\[-1.0ex] 
  Cohort II & 0.58 & 0.81 & 0.81 & 0.52 & 0.86 & 0.46 & 0.44 & 0.57 & 0.12 \\[-1.5ex]
  
      & (0.017) & (0.017) & (0.017) & (0.022) & (0.018) & (0.019) & (0.009) & (0.009) & (0.005) \\[-1.0ex] 
  Cohort III & 0.64 & 0.87 & 0.89 & 0.58 & 0.91 & 0.51 & 0.47 & 0.57 & 0.10 \\[-1.5ex]

       & (0.025) & (0.022) & (0.023) & (0.033) & (0.023) & (0.031) & (0.015) & (0.015) & (0.011) \\ 
   \hline
   \hline
\end{tabular}
\begin{tablenotes}
	\item [] \footnotesize{\textit{Notes:} Standard errors of estimates, reported between brackets, are calculated by 1000 bootstraps on the observed occupational transitions used to estimate $\mathbf{P}$.}
	\item[] \footnotesize{\textit{Source:} Our calculations based on Survey on Household Income and Wealth (Bank of Italy).}
\end{tablenotes}
\end{threeparttable}
\label{tab:estimatedParameters}
\end{table}

For children in Wc incentives to change their class  have been decreasing over time ($\lambda_M$ rose from 0.44 to 0.64), while opportunities have been expanding ($\theta^{\max}$ jumped from 0.66 to 0.89). Therefore, the increased overall persistence in Wc (from 67\% to 72\%, see Table \ref{tab:Transitionprobabilities_estimate}) covers two opposite phenomena, with the impact of declining incentives prevailing on the higher opportunities.
For children in Uc the incentives to change their class have been increasing over time ($\lambda_U$ rose from 0.66 to 0.87), and opportunities to fall behind have been reducing ($\theta^{\min}$ climbed from 0.38 to 0.58). As a consequence, the decreased overall persistence in Uc (from 55\% to 31\%, see Table \ref{tab:Transitionprobabilities_estimate}) again covers two opposite phenomena, with the impact of increasing incentives to move prevailing on the lower (opportunities) possibilities of sliding down.
Finally, for children in Mc, as the ones in Wc, opportunities to move upward have been increasing  ($\theta^{\max}_M$ increased from 0.71 to 0.91), while opportunities to move downward have been declining ($\theta^{\min}_M$ increased from 0.33 to 0.51); income incentives have been working in the same directions, i.e. the increased $\lambda_M$ favoured the outflow toward Wc, while the increase in $\lambda_U$ discouraged any upward movement toward Uc. The slightly declined persistence over time is the result of these competing forces.

Overall, equality of opportunity as measured by $I_{OPP}$ calculated by Eq. (\ref{eq:indexEqualityOpportunity}) are slightly increased from Cohort I to Cohort II (from 0.55 to 0.57), but steady from Cohort II to Cohort III. However, from the estimated parameters we know that the outcome is the sum of increasing opportunities to move upward (both $\theta^{\max}$ and $\theta^{\max}_M$ increased over time) and declining opportunities to move downward (both $\theta^{\min}$ and $\theta^{\min}_M$ increased as well). Finally, income incentives as measured by $I_{LOI}$ calculated by Eq. (\ref{eq:indexLackOfIncentive}) shows a positive trend (from 0.15 to 0.10) as the outcome of opposing forces, with the increased income incentives to move downward for children with fathers in Mc and Uc, and the lower income incentives to move upward from Wc and Mc. From another perspective, the increased importance of income incentives can be appreciated by observing that in Cohort I children exploited 0.40/0.55 = 73\% of opportunities, while in Cohort III this value raised to 0.47/0.57 = 82\%.

\subsubsection{The Determinants of Income Incentives \label{sec:determinantIncomeIncentives}}

From the policy perspective it would be crucial to identify the contribution of (occupational) \textit{return premium}, \textit{risk premium} and the \textit{cost of access} (see Eqq. (\ref{eq:ConditionIncomeIncentives}) and (\ref{eq:ConditionIncomeIncentivesUc})), to the increasing trend in $\lambda_M$ and $\lambda_U$ reported in Table \ref{tab:estimatedParameters}. Unfortunately, as we will discuss below, the available information only allows for a limited investigation of this issue.

Assume that the log of income can be take as proxy for the utility of children. Moreover, assume that, at the moment in which children took their occupational decision, their expected utility and its standard deviation can be proxied by the observed mean and standard deviation of the distribution of the log of income of their working life, i.e. income differentials among occupations are assumed to be steady along working life.\footnote{We consider as individual income the net total available income of individuals, coded by label Y in SHIW, expressed in constant prices by the consumer price index with base year 2013 provided by ISTAT.} Table \ref{tab:averageAndStandardDeviationRelativeNetIncome} reports the ratios between sample means (proxy for return premium) and standard deviations (proxy for risk premium) of the log of incomes for Cohorts I, II and III for the period 1995-2012.\footnote{In particular, we calculate the mean and standard deviation of the distribution of the log of incomes for each wave and for each cohort. We then calculate their weighted mean over all nine waves, with weights proportional to the number of observations in each wave, and finally we calculate the ratios. Since the values reported in Table \ref{tab:averageAndStandardDeviationRelativeNetIncome} refer to the ratio between means of the log of incomes, the ratio between the means of incomes are much more higher; for example, the ratio of means of the log of income equal to 1.055 of Mc with respect to Wc in Cohort I implies that to an income of 1000 in Wc would correspond to an income of $1000^{1.055}=1462.18$ in Mc, i.e. 46.2\% higher and not just 5.5\%.} 

\begin{table}[htbp]
\centering
\caption{\small {\bf Ratios between sample means (proxy for return premium) and standard deviations (proxy for risk premium) of the log of incomes for Cohorts I, II and III.}} 
\begin{threeparttable}
\begin{tabular}{l|rrr}
  \hline
  \hline
Return premium      & $\mu_M$/$\mu_W$ & $\mu_U$/$\mu_M$ \\ 
  \hline
Cohort I   & 1.055        & 1.049 \\ 
Cohort II  & 1.050 		  & 1.041 \\ 
Cohort III & 1.048 		  & 1.026 \\ 
   \hline
Risk premium & $\sigma^2_M$/$\sigma^2_W$ & $\sigma^2_U$/$\sigma^2_M$ \\ 
  \hline
Cohort I  & 0.480 & 1.105 \\ 
Cohort II  & 0.486 & 1.088 \\ 
Cohort III & 0.832 & 0.783  \\ 
     \hline
   \hline
\end{tabular}
\begin{tablenotes}
\item[] \footnotesize{\textit{Source:} Our calculations based on Survey on Household Income and Wealth (Bank of Italy).}
\end{tablenotes}
\end{threeparttable}
\label{tab:averageAndStandardDeviationRelativeNetIncome}		
\end{table}

As expected, on average incomes are higher in Uc and lower in Wc, and therefore Condition (\ref{eq:assumptionDifferentExpectedUtilityDifferentOccClass}) is respected. Instead, the picture of the standard deviation of the distribution of incomes is more complex: in the all three cohorts Wc displays the highest values (essentially caused by the highest unemployment rate in Wc), while Uc has higher values than Mc for the first two cohorts, while the opposite holds for Cohort III. Overall, figures reported in Table \ref{tab:averageAndStandardDeviationRelativeNetIncome} support Assumption \ref{assumption:lambda_s}, excluding the higher value of standard deviation in Mc than Uc in Cohort III.

The decline in the return premium of belonging to Mc instead of Wc (i.e. $\mu_M$/$\mu_W$) and the increase in risk premium $\sigma^2_M$/$\sigma^2_W$ can contribute to explain the estimated increase in $\lambda_M$. In the same manner, the decline in the return premium of belonging to Uc instead of Mc (i.e. $\mu_U$/$\mu_M$) can contribute to explain the estimated increase in $\lambda_U$; however, the decline in risk premium $\sigma^2_U$/$\sigma^2_M$ tends to push upward $\lambda_U$.

Our model points to an increasing cost of meeting the required skill to access to Mc $c^e_M$ and Uc $c^e_U$ as other explanatory causes of the upward trend of $\lambda_M$ and $\lambda_U$. Unfortunately, we have no specific information on the level of tuition fees and financial supports to Italian students, the most likely determinants of $c^e_M$ and $c^e_U$. In particular, we have no data for tertiary education, which is the most likely type of education to access to Uc (see, e.g., \citealp{Checchi1997} for Italy and \cite{HauserEtAl2000} for US). There exists however a common view in literature that from the end of 1980's the tuition fees in Italian universities started increasing, while the restrictions on public budget curbed the financial support to students. At the beginning of 2000's \citet{OECD2006} ranks Italy at the ninth highest position for tuition fees' levels among European countries, and it also points to the very low percentage of Italian students with a scholarship. For the same period, \citet{Istat2005} confirms that in Italy the access to tertiary education is checked by the high and increasing level of tuition fees, and highlights the growing accommodation cost for students.

\section{Concluding Remarks \label{sec:ConcludingRemarks}}

\citet{Corak2013} discusses how the observed downward trend in mobility in most of western countries is generally perceived as a source of declining efficiency and justice (\Citealp{Gilbert2011} and \citealp{Stiglitz2012}).
On the contrary, we have documented that Italian children born in the period 1951-1965 experienced a growing occupational mobility with respect to children born in the period 1940-1951, while those born after 1965 no changes in mobility. However, the picture is still gloomy because this steady mobility is unfortunately largely ascribable to higher downward mobility from upper-middle occupational classes and to the fading structural mobility in favour of upper-middle occupational classes present just after the II World War in Italy. We have also found that equality of opportunities are far from the perfection and steady for those born after 1965. The change in income incentives instead played a major role, favouring higher downward mobility from upper-middle occupational classes and lower upward mobility from lower class. These estimated trend in income incentives could be (partially) explained by the observed trend in differential returns and riskiness of occupations, together with anecdotal evidence of the increasing cost of education in Italy.

Our analysis contains two main theoretical limitations. First, we have assumed that individuals know their ability (proxied by $\theta$ in the model) when they take their decision of occupation. However, the uncertainty on the level of own ability can induce risk-adverse individuals not to educate, in particular if they poor.
Second, we have adopted a definition of equality of opportunity strictly related to the conception of ``level the playing field''; \citet{Roemer1998} advances another conception which he calls the ``nondiscrimination or merit principle''. With his words, it holds when ``in the competition for the position in society, all individuals who possess the attributes relevant for the performance of duties of the position in question should be included in the pool of eligible candidates, and that an individual's possible occupancy of the position should be judged only with respect to those relevant attributes'' (\citealp[p.17]{Roemer2000}). For example, race and sex should not affect the competition for a position, but in the real world it is likely to be so. In this regard another limitation of our model is its inability to identify the individual contribution to social (im)mobility of genes, cultural traits and family background, which is the subject of an intense debate in literature (see, e.g., \citealp{arrowEtAl2000} and \citealp{HeckmanMosso2014}).

Future research should aim to firstly tackle the two limitations discussed above, with the most promising directions of research regarding the impact of credit market imperfections on socio-economic mobility, and the identification of the contribution (if any) to observed (im)mobility of the transmission of cultural traits and family background as opposed to genes (see, e.g., \citealp{BjorklundJantti2009}). Finally, our model could be extended to investigate a possible (negative) relationship between inequality and intergenerational mobility (the so-called ``Great Gatsby Curve''), which a recent empirical literature has found in US data (\citealp{Krueger2012} and \citealp{Corak2013}).

\end{document}